\documentclass[12pt]{iopart}

\usepackage{graphicx}
\usepackage{iopams}
\usepackage[pdfborder=000, colorlinks, linkcolor=blue, citecolor=blue]{hyperref}
\usepackage[numbers,sort&compress]{natbib}

\begin{document}

\title[]{Homophily on social networks changes evolutionary advantage in competitive information diffusion}

\author{Longzhao Liu$^{1,4}$, Xin Wang$^{1,5}$, Yi Zheng$^{1}$, Wenyi Fang$^{2,3}$, Shaoting Tang$^{1,3,6}$ and Zhiming Zheng$^{1,3}$}

\address{$^1$ LMIB, NLSDE, BDBC and School of Mathematics and Systems Science, Beihang University, Beijing 100191, China}
\address{$^2$ School of Mathematical Sciences, Peking University, Beijing, 100871, China}
\address{$^3$ PengCheng Laboratory, Shenzhen, 518055 , China}
\address{$^4$ ShenYuan Honors College, Beihang University, Beijing 100191, China}
\address{$^5$ Department of Mathematics, Dartmouth College, Hanover, NH 03755, USA}
\address{$^6$ Author to whom any correspondence should be addressed} 

\ead{tangshaoting@buaa.edu.cn}

\vspace{10pt}

\begin{abstract}
Competitive information diffusion on large-scale social networks reveals fundamental characteristics of rumor contagions and has profound influence on public opinion formation. There has been growing interest in exploring dynamical mechanisms of the competing evolutions recently. Nevertheless, the impacts of population homophily, which determines powerful collective human behaviors, remains unclear. In this paper, we incorporate homophily effects into a modified competitive ignorant-spreader-ignorant (SIS) rumor diffusion model with generalized population preference. Using microscopic Markov chain approach, we first derive the phase diagram of competing diffusion results and examine how competitive information spreads and evolves on social networks. We then explore the detailed effects of homophily, which is modeled by a rewiring mechanism. Results show that homophily promotes the formation of divided ``echo chambers" and protects the disadvantaged information from extinction, which further changes or even reverses the evolutionary advantage, i.e., the difference of final proportions of the competitive information. We highlight the conclusion that the reversals may happen only when the initially disadvantaged information has stronger transmission ability, owning diffusion advantage over the other one. Our framework provides profound insight into competing dynamics with population homophily, which may pave ways for further controlling misinformation and guiding public belief systems. Moreover, the reversing condition sheds light on designing effective competing strategies in many real scenarios.
\end{abstract}

\maketitle
%
%

\section{Introduction}
Owing to the rapid development of Internet technologies, the dynamical evolutions of information diffusion on large-scale social networks have been widely concerned and studied in recent years \cite{castellano2009statistical,chang2018co,pei2014searching,zhang2016dynamics,gomez2013diffusion,li2014rumor,vosoughi2018spread,kermack1927contribution,watts2002simple,liu2015events}. How to calculate the final scope of information delivery? How to predict the threshold of information dissemination? What's the coupled effects of information diffusion and other evolution processes on multiple networks? Exploring the underlying dynamical mechanisms of information diffusion is of vital significance for understanding these important questions, which can further instruct powerful applications in many different fields, such as controlling rumor spreading, promoting innovations, designing efficient marketing strategies and etc \cite{li2015multiple,pei2013spreading,torok2017cascading}.

Of particular interest, competitive information diffusion on social networks, which reproduces the ubiquitous situations where individuals are exposed to multiple polarized information related to the same social event, has attracted great attention recently \cite{alon2010note,liu2016shir, crokidakis2013role}. Understanding the diffusion patterns of competitive information helps to deal with some great challenges nowadays. A direct application scenario is to understand the simultaneous spreading of the truth and the rumors \cite{zhang2014dynamic, friedkin2017truth}. This is of great significance not only in physics, but also in economics and social science, considering the fact that misinformation has been recorded as one of the main threats to human society by World Economic Forum (WEF) \cite{del2016spreading}. Moreover, the final competing diffusion results have profound influence on the formation of public opinions \cite{ben2011exact,liggett2013stochastic,nowak1990private, lambiotte2008dynamics, shao2009dynamic,trpevski2014discrete}. A typical example is the US presidential election, where the outcome of competitive information diffusion on large-scale social networks, like Facebook and Twitter, would directly affect the election result \cite{bovet2019influence,grinberg2019fake}.

The competing dynamics was first studied on epidemic spreading where two competitive diseases infect the same population \cite{funk2010interacting, sahneh2014competitive, sanz2014dynamics}. Newman utilized the generalized susceptible-infective-removed (SIR) model to describe the transmission of two pathogens on networks \cite{newman2005threshold}. Karrer {\it et al.} further studied the dynamics of competing diseases with cross immunity and derived the theoretical phase diagram of the system \cite{karrer2011competing}. Leventhal {\it et al.} applied competitive susceptible-infected-susceptible (SIS) model to describe the evolution of infectious diseases and highlighted the significant role of network heterogeneity \cite{leventhal2015evolution}.

Inspired by the idea that the dynamical process of information diffusion is analogous to epidemic spreading to some extent, most literatures attempted to use epidemic-like model to describe the competitive information diffusion \cite{zhang2013rumor, stanoev2014modeling, jie2016study}. Trpevski {\it et al.} proposed competitive SIS model with stubbornness and completely asymmetry preference, where the competing rumors satisfy cross immunity and individuals always select rumor $1$ when they are informed of two rumors simultaneously \cite{trpevski2010model}. Furthermore, some studies tried to incorporate the unique consumption patterns of information diffusion compared to the disease spreading, which makes the model closer to reality. Wang {\it et al.} proposed a theoretical framework that considered neighborhood influences and found rich dynamics of two competing ideas \cite{wang2012dynamics}.

Despite the efforts and progress, it remains largely unknown how human behaviors, especially those collective evolutions caused by the group psychology, influence the competitive information diffusion. Previous studies have found that population homophily, which means individuals prefer to seek for like-minded people while avoid further communications with those who hold opposite opinions, is a widespread characteristic of large-scale social networks \cite{boutyline2017social, flaxman2016filter, aiello2012friendship}. In particular, homophily was verified as the main psychological reason for the formation of echo chambers \cite{colleoni2014echo}. This indicates that homophily is an essential factor of collective behaviors in competitive information diffusion processes, which not only influences the efficiency of information spreading, but also changes the potential diffusion paths, i.e., affects the underlying network topology. However, there still lacks a proper understanding of the detailed impacts of homophily on diffusion results of competitive information.

To fill this theoretical gap, in this paper, we propose a theoretical framework which incorporates population homophily into a modified competitive SIS model with generalized population preference \cite{granell2014competing}. Both theoretical analysis and simulation results are provided to show how competitive information diffuses and evolves on social networks. We find that homophily can significantly change and even reverse the evolutionary advantage, i.e., the difference of final proportions of the two information. However, the reversals only happen when the initially disadvantaged information has stronger transmission ability but loses population preference. This indicates that to win the competing evolutions, the best way for the information is to take the diffusion advantage regardless of population preference. Our work reveals how population homophily influences the diffusion processes and the evolutionary results of competitive information, which provides profound insight into misinformation spreading as well as public opinion formation on large-scale social networks.

\section{Model description}\label{model}
Consider an undirected network with $N$ nodes, whose adjacent matrix is denoted as $A=(a_{ij})_{N\times N}$. If there exists an edge between node $i$ and node $j$, $a_{ij}=1$. Otherwise, $a_{ij}=0$. We adopt ignorant-spreader-ignorant (SIS) rumor spreading model to describe the diffusion process of a single piece of information, where ignorant (I) represents an individual who has not known the information and spreaders (S) stand for the individuals who are able to spread information to its neighbors, corresponding to the susceptible and infected populations in classical epidemic model, respectively \cite{pei2015exploring} . An important characteristic of the SIS model is the reinfection mechanism, i.e. the recovered nodes can be infected again, which could effectively describe the multi-round information dissemination. In addition, the reinfection mechanism provides a simple and natural way to characterize the viewpoint changing process in multiple information diffusion circumstances.

In this work, we focus on the situation where two pieces of competitive information, denoted as information $1$ and information $2$, spread simultaneously. Complying with previous studies, we assume that information $1$ and information $2$, such as the truth and the rumors, are competitive with exclusiveness \cite{souza2012dynamical}. This means that all individuals could only support one piece of information at any time. Thus, the population could be divided into three classes according to their states: ignorant ($I$), spreader of information $1$ ($S_{1}$), spreader of information $2$ ($S_{2}$). The ignorant has not learned any information or is confused about which information to support. $S_1$ strongly supports information $1$ and has a probability $\lambda_1$ to spread information $1$ to its ignorant neighbors. Meanwhile, $S_1$ forgets the information or becomes confused again because of the self-awareness or the external social influence, i.e., the individual changes its state from $S_1$ to $I$, with probability $\mu_1$ \cite{wang2017promoting}. Similarly, $S_{2}$ transmits information $2$ to its neighbors with probability $\lambda_2$ and becomes ignorant with probability $\mu_2$. In addition, we assume that all individuals are stubborn once they make a choice, which means that the spreaders of one information can not be persuaded by their neighbors to support the other one directly \cite{ghaderi2014opinion}.

It is worthy of note that there exists a special and important situation where the ignorant is informed of two competitive information at the same time. Previous works studied the scenario where information $1$ always has a higher priority \cite{trpevski2010model}. To describe the population preference in a general way, here we denote a new parameter $\alpha$ $(0\leq\alpha\leq 1)$ as the probability that the ignorant will choose to support and spread information $1$ when receives both information simultaneously.

Moreover, we incorporate rewiring mechanism into our model to describe the wide-existing homophily on large-scale social networks, which is also well-known as ``echo chamber" phenomenon. At each time step, the links between $S_1$ and $S_2$ break with probability $p$. Meanwhile, new links will generate from one of the broken links' endpoints to the randomly selected individuals who support the same information or in  ignorant states. In this way, we mimic the homophily behaviors that individuals tend to avoid further communications with those who hold opposite opinions while prefer to seek for those who support the same information. The rewiring probability $p$ reflects the strength of homophily among populations.

\begin{figure}
\begin{indented}
\item[]\includegraphics[width=13cm]{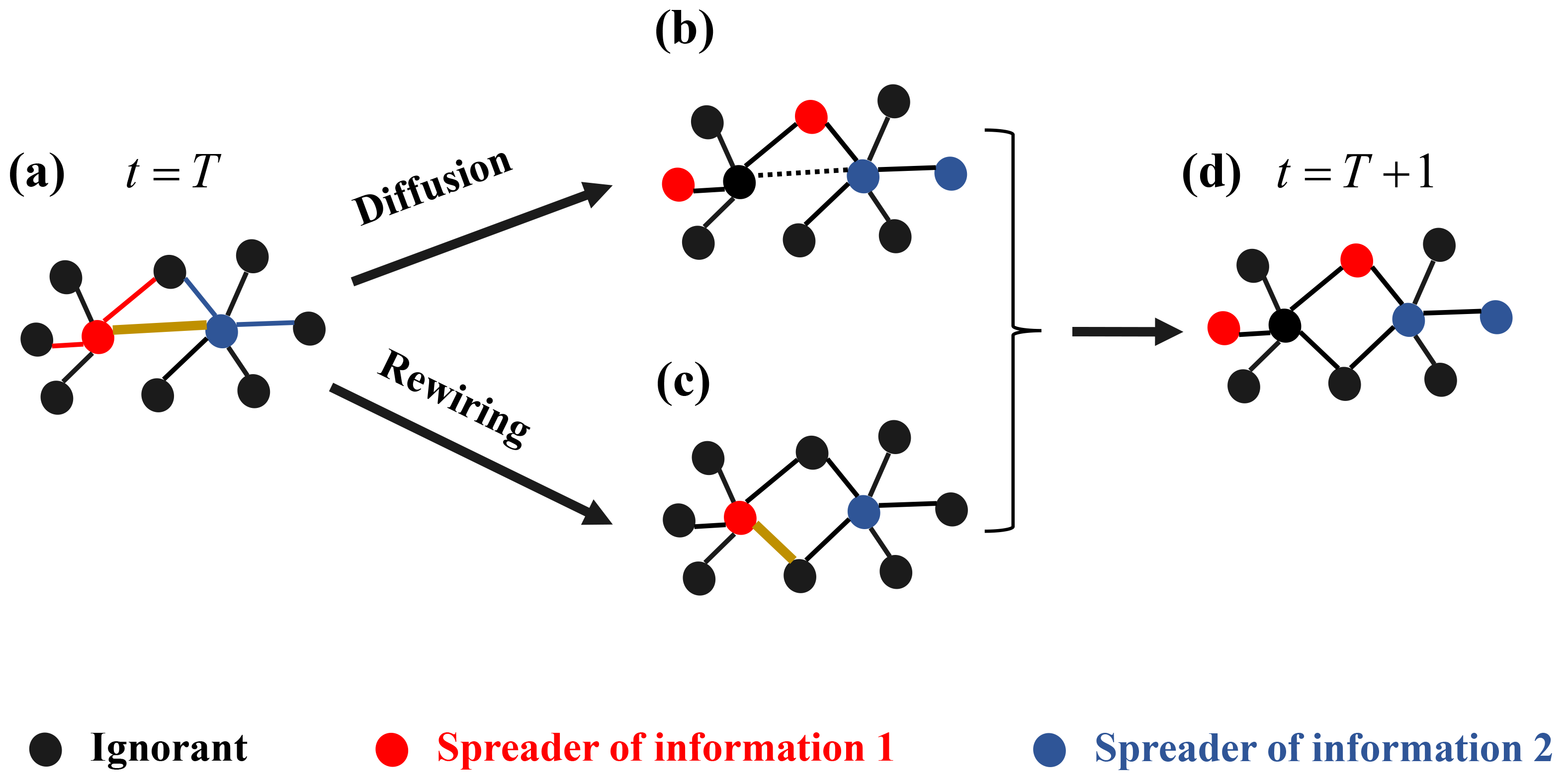}
\end{indented}
\caption{Dynamical model of competitive information diffusion on social networks. Nodes represent the individuals and edges represent the connections between them. (a) Network structure and individuals' states at time $T$. Ignorants, spreaders of information $1$ ($S_1$) and spreaders of information $2$ ($S_2$) are represented by black, red and blue nodes, respectively. (a)(b) Diffusion process. $S_1$ spreads information $1$ to its neighbors along red edges while $S_2$ diffuses information $2$ along blue edges. If an ignorant individual receives both information simultaneously, he/she will choose to support information $1$ with probability $\alpha$. Meanwhile, spreaders have a probability of becoming ignorants again. (a)(c) Rewiring process. The link between $S_1$ and $S_2$ breaks with probability $p$ and a new link will generate from one of the spreaders to a randomly chosen individual who is ignorant or support the same information, as shown by the yellow edges. (d) Network structure and individuals' states at time $T+1$. It is composed of the nodes' state in (b) and the network structure in (c).}
\label{sketch}
\end{figure}

In summary, our model is composed of two dynamical processes: diffusion process and rewiring process, as shown in figure \ref{sketch}:
\begin{enumerate}
	\item\textbf{Diffusion process.} At each time step, $S_1$ and $S_2$ transmit information to ignorant neighbors with probability $\lambda_{1}$ and $\lambda_{2}$, respectively. Meanwhile, $S_1$ and $S_2$ become ignorant again with probability  $\mu_{1}$ and $\mu_{2}$. Specially, if an ignorant receives both competitive information simultaneously, it chooses to become $S_1$ with probability $\alpha$. Otherwise, with probability $1-\alpha$ it becomes $S_{2}$.
	\item\textbf{Rewiring process.} At each time step, the links between $S_1$ and $S_2$ break with probability $p$. Meanwhile, new links will generate, which connect one of the endpoints of the broken links to the randomly chosen individuals who are ignorants or spread the same information.
\end{enumerate}
\section{Theoretical framework}
\subsection{Modified competitive SIS model with population preference}

We first explore how the modified competitive SIS model without homophily behaves on networks. Let $\widetilde{q_{i}}^{S_1}(t)$ denotes the probability that node $i$ changes its state from $I$ to $S_1$ at time $t$. Based on the description of diffusion process in Section \ref{model},  $\widetilde{q_{i}}^{S_1}(t)$ can be written as
\begin{equation}
\widetilde{q_{i}}^{S_1}(t)=(1-q_i^{S_1}(t))q_i^{S_2}(t)+\alpha(1-q_i^{S_1}(t))(1-q_i^{S_2}(t)),
\label{1}
\end{equation}
where $q_i^{S_1}(t)$ and $q_i^{S_2}(t)$ indicate the probabilities that node $i$ is not informed of information $1$, not informed of information $2$ if $i$ is an ignorant at time $t$, respectively. Thus, the first term of equation (\ref{1}) represents the probability that node $i$ is informed of information $1$ but not informed of information $2$, while the second term expresses the probability that node $i$ receives both information concurrently and choose to spread information $1$.

Let $p_i^X(t)$ denotes the probability that node $i$ is in $X$ state at time $t$, $X\in \{I, S_1, S_2\}$. Clearly, $p_i^X(t)$ satisfies $p_i^I(t)+p_i^{S_1}(t)+p_i^{S_2}(t)=1$. Then we have
\begin{equation}
\eqalign{
q_i^{S_1}(t)&=\prod\limits_{i\neq j} (1-\lambda_1a_{ij}p_j^{S_1}(t))\\
q_i^{S_2}(t)&=\prod\limits_{i\neq j} (1-\lambda_2a_{ij}p_j^{S_2}(t)).}
\label{3}
\end{equation}

Similarly, the probability that node $i$ in ignorant state becomes a spreader of information $2$ at time $t$, denoted as $\widetilde{q_{i}}^{S_2}(t)$, is
\begin{equation}
\widetilde{q_{i}}^{S_2}(t)=q_i^{S_1}(t)(1-q_i^{S_2}(t))+(1-\alpha)(1-q_i^{S_1}(t))(1-q_i^{S_2}(t)).
\label{2}
\end{equation}

Therefore, the evolutionary equations of the competitive SIS model with population preference can be derived by microscopic Markov chain approach, which read as
\begin{equation}
\eqalign{
p_i^I(t+1)&=p_i^I(t)q_i^{S_1}(t)q_i^{S_2}(t)+p_i^{S_1}(t)\mu_1\\
&+p_i^{S_2}(t)\mu_2\\
p_i^{S_1}(t+1)&=p_i^I(t)\widetilde{q_{i}}^{S_1}(t)+p_i^{S_1}(t)(1-\mu_1)\\
p_i^{S_2}(t+1)&=p_i^I(t)\widetilde{q_{i}}^{S_2}(t)+p_i^{S_2}(t)(1-\mu_2).
}
\label{evolution equation}
\end{equation}


\subsection{Competitive information diffusion with homophily}

We then examine the dynamics of competitive information diffusion with homophily, described by the rewiring process in our model. Note that under this condition, the structure of underlying network evolves over time and is dependent on the current distribution of the population states. On the other hand, the diffusion process also relies on the changing topology of the network. Therefore, we have to consider the coupled evolutions of the nodes' states and the network structure. Denote $f_t(A,\boldsymbol x)$ as the probability distribution that $A(t)=A$ and $\boldsymbol x(t)=\boldsymbol x$ in the dynamical system. Here $A(t)=(a_{ij}(t))_{N\times N}$ is the adjacent matrix and $\boldsymbol x(t)=(x_1(t), x_2(t), ... ..., x_N(t))$ is the state vector of all nodes at time $t$. $x_i(t)=0, 1, 2$ represents that node $i$ is in state $I$, $S_1$ and $S_2$ at time $t$, respectively.

In this section, we give mathematical description of evolution process from time $t$ to $t+1$, i.e. calculating $f_{t+1}(B,\boldsymbol y)$ for any possible $A(t+1)=B=(b_{ij})_{N\times N}$ and $\boldsymbol x(t+1)=\boldsymbol y=(y_1, y_2, ... ..., y_N)$.

Firstly, given a certain system state $A(t)=A$ and $\boldsymbol x(t)=\boldsymbol x$ at time $t$. The probability that ignorant node $i$ is not informed of information 1 or information 2 can be written as
\begin{equation}
\eqalign{
q_i^{S_1}(t)&=\prod\limits_{i\neq j} (1-\lambda_1a_{ij}(t)\delta(x_j(t)-1))\\
q_i^{S_2}(t)&=\prod\limits_{i\neq j} (1-\lambda_2a_{ij}(t)\delta(x_j(t)-2)),
}
\label{4}
\end{equation}
where $\delta(x)=1$ if $x=0$ and $\delta(x)=0$ if $x\neq0$. Then we can derive the probabilities that node $i$ changes its state from $I$ to $S_1$ and $S_2$, i.e., $\widetilde{q_{i}}^{S_1}(t)$ and $\widetilde{q_{i}}^{S_2}(t)$, by substituting equation (\ref{4}) into equations (\ref{1}) and (\ref{2}). Thus, the evolutionary equations of population states read
\begin{equation}
\eqalign{
p_i^I(t+1)&=\delta(x_i(t))q_i^{S_1}(t)q_i^{S_2}(t)+\delta(x_i(t)-1)\mu_1\\
&+\delta(x_i(t)-2)\mu_2\\
p_i^{S_1}(t+1)&=\delta(x_i(t))\widetilde{q_{i}}^{S_1}(t)+\delta(x_i(t)-1)(1-\mu_1)\\
p_i^{S_2}(t+1)&=\delta(x_i(t))\widetilde{q_{i}}^{S_2}(t)+\delta(x_i(t)-2)(1-\mu_2).
}
\label{5}
\end{equation}

Further, the conditional probability that node $i$ is in state $y_i$ at time $t+1$, which is denotes as $p^{y_i}$, can be written as
\begin{equation}
p^{y_i}=\delta(y_i)p_i^I(t+1)+\delta(y_i-1)p_i^{S_1}(t+1)+\delta(y_i-2)p_i^{S_2}(t+1).
\label{6}
\end{equation}
Therefore, the probability that nodes' states evolve from $\boldsymbol x$ to $\boldsymbol y$ is
\begin{equation}
p_{\boldsymbol x \to \boldsymbol y}=\prod\limits_{i=1}^{N}p^{y_i}.
\label{7}
\end{equation}

Let $p_{A \to B}$ denote the probability that network structure evolves from $A$ to $B$. The number of links between $S_1$ and $S_2$ at time $t$ is
\begin{equation}
L= \frac{1}{2}\sum\limits_{(i, j)} \delta(x_i(t) x_j(t) a_{ij}(t)-2).
\label{8}
\end{equation}
At time $t+1$, the number of remaining $S_1 S_2$ links which is not broken at time $t$ is
\begin{equation}
L^*= \frac{1}{2}\sum\limits_{(i, j)} \delta(x_i(t) x_j(t) b_{ij}-2).
\label{8}
\end{equation}
According to the rewiring mechanism defined in Section \ref{model}, we have
\begin{equation}
p_{A \to B}=(1-p)^{L^*} \left[\frac{p}{2(I(t)+S_1(t))}\right]^{l_1}\left[\frac{p}{2(I(t)+S_2(t))}\right]^{L-L^*-l_1},
\label{9}
\end{equation}
where $l_1=\frac{1}{2}\sum\limits_{(i,j)}\phi(b_{ij}-a_{ij}(t))$ for all $(i,j)$ that satisfies $\{(i,j)|x_i(t)=1\; or \; x_j(t)=1\}$, and $\phi(x)=1$ if $x=1$, otherwise $\phi(x)=0$. Here $l_1$ calculates the number of new links rewiring from the $S_1$ endpoints, i.e. new $S_1S_1$ and $S_1I$ links at time $t$. Hence, the first formula $(1-p)^{L^*}$ is the conditional probability that there are $L^*$ remaining $S_1 S_2$ links at time $t$, and the second formula $\left[p/(2I(t)+2S_1(t))\right]^{l_1}*\left[p/(2I(t)+2S_2(t))\right]^{L-L^*-l_1}$ represents the conditional probability that the other $L-L^*$ broken $S_1 S_2$ links rewire, among which $l_1$ new links rewire from the $S_1$ endpoints. Note that the second formula here is actually an approximation which does not take the multiple edges or loops into consideration, since the real social networks are large and sparse.

Finally, note that the evolution probabilities of nodes' states and network structure from time $t$ to $t+1$ are independent. Therefore, summing up all the possible system states at time $t$, the coupled probability distribution $f_{t+1}(B,\boldsymbol y)$ at time $t+1$ can be calculated by
\begin{equation}
\fl\eqalign{
&f_{t+1}(B,\boldsymbol y)\\
&=\sum\limits_{(A,\boldsymbol x)}f_{t}(A, \boldsymbol x)*p_{\boldsymbol x \to \boldsymbol y}*p_{A \to B}\\
&=\sum\limits_{(A,\boldsymbol x)}f_{t}(A, \boldsymbol x)*\prod\limits_{i=1}^{N}p^{y_i}*(1-p)^{L^*}*\left[\frac{p}{2(I(t)+S_1(t))}\right]^{l_1}\left[\frac{p}{2(I(t)+S_2(t))}\right]^{L-L^*-l_1}.
}
\end{equation}
\section{Results}
We start from the Erd\"{o}s-R\'{e}nyi (ER) random graphs with $N=1000$ nodes. The average degree is $\langle k \rangle=10$ and initially ten $S_1$ and ten $S_2$ are randomly selected. To eliminate the fluctuation, the numerical simulation results in the rest of the paper is the average of 100 times. The stable states of the dynamical system is approximated by the simulations running for $500$ steps. Additionally, for simplicity and without loss of generality, we set $\mu_1=\mu_{2}=\mu$.

\subsection{The effects of diffusion advantage and population preference}
Firstly, we study how diffusion advantage and population preference affect the competing diffusion results. Here the diffusion advantage refers to the difference between transmission probabilities of the competitive information. We set $p=0$ which excludes the influence of homophily.

\begin{figure}
\begin{indented}
\item[]\includegraphics[width=13cm]{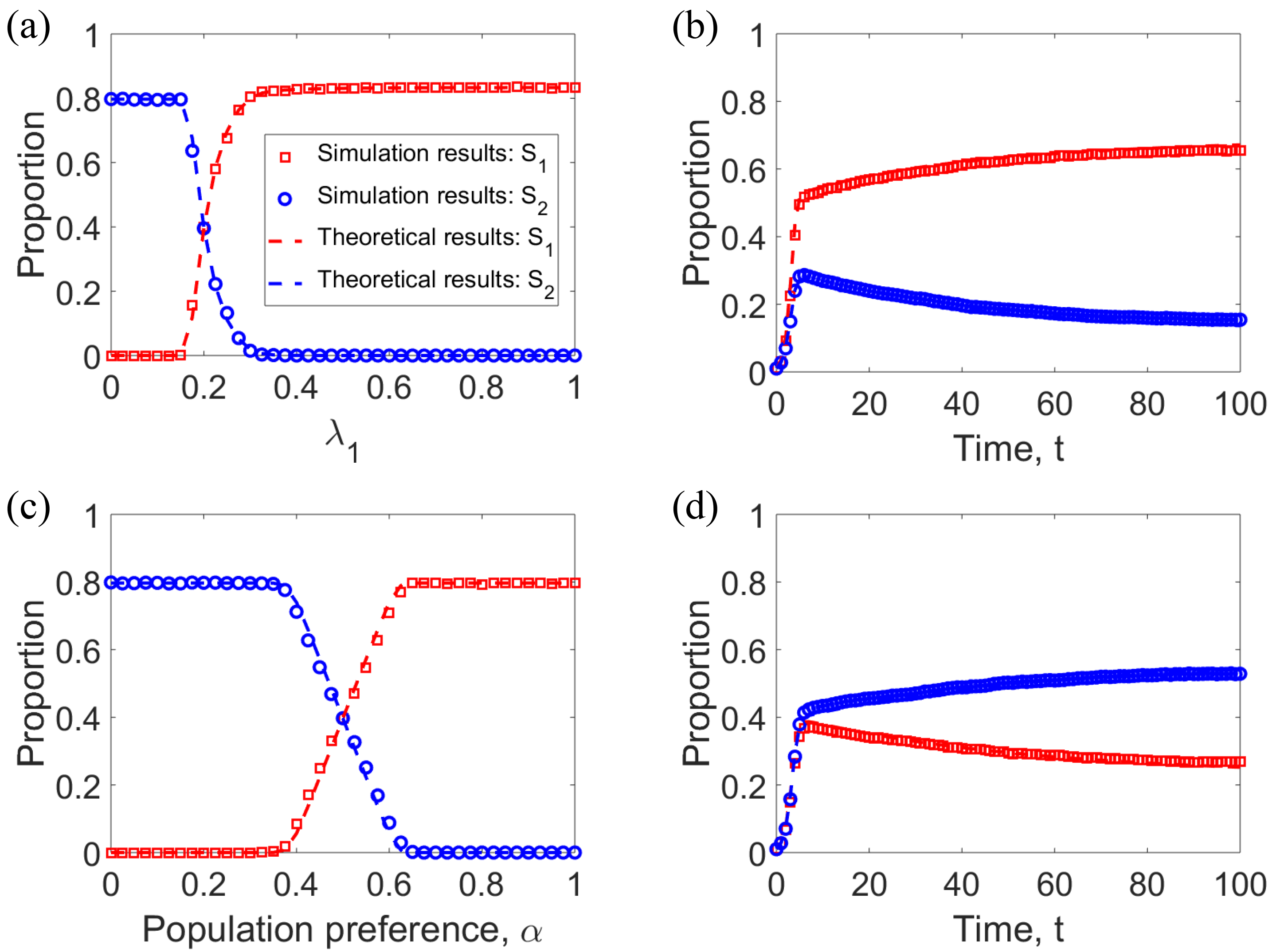}
\end{indented}
\caption{The effects of diffusion advantage and population preference on competing diffusion results. Theoretical predictions provided by numerical solutions of equations (\ref{1})- (\ref{evolution equation}) are shown by dash lines. Note that the rewiring probability $p$ is set to be $0$, which excludes the influence of homophily. (a) How the diffusion advantage, i.e. the difference between transmission probability, affects competitive information diffusion. We fix $\lambda_2=0.2$, $\mu=0.2$, $\alpha=0.5$ and change $\lambda_1$ from 0 to 1. (b) Time evolutions of competing results when diffusion advantage exists. The parameters are as follows: $\lambda_1=0.25$, $\lambda_2=0.2$, $\mu=0.2$, $\alpha=0.5$. (c) The impact of population preference. We set $\lambda_1=\lambda_2=0.2$, $\mu=0.2$. The competing results are presented as a function of $\alpha$. (d) Time evolutions of competing results when population preference exists. We set $\lambda_1=\lambda_2=0.2$, $\mu=0.2$, $\alpha=0.45$.}
\label{population preference}
\end{figure}

In figure \ref{population preference}(a), we present the effect of diffusion advantage. To eliminate the influence of population preference, $\alpha$ is set to be $0.5$. We fix $\lambda_2=0.2, \mu=0.2$ and change $\lambda_1$ from $0$ to $1$. Results show that the competing results experience three stages as $\lambda_1$ increases. Initially when $\lambda_1 \leq 0.16$, $S_1$ becomes extinct in the competition. When $0.16 < \lambda_1 < 0.3$, the proportion of $S_1$ increases sharply and $S_1$, $S_2$ coexist. Finally when $\lambda_1 \geq 0.3$, only $S_1$ survives. Figure \ref{population preference}(b) shows detailed time evolutions of competing results when information $1$ has the diffusion advantage and $S_1$, $S_2$ coexist. In figure \ref{population preference}(c), we provide the impact of population preference. We fix $\lambda_1=\lambda_2=0.2$ and change $\alpha$ from 0 to 1. Similarly, $S_1$ first vanishes, then rises rapidly, and dominants the population in the end as $\alpha$ increases. A typical example for the time evolutions of competing results when population preference exists is shown in figure \ref{population preference}(d). Note that in all subfigures, our theoretical predictions provided by numerical solutions of equations (\ref{1})- (\ref{evolution equation}) agree well with the simulation results, as shown by dash lines.

\begin{figure}
\begin{indented}
\item[]\includegraphics[width=13cm]{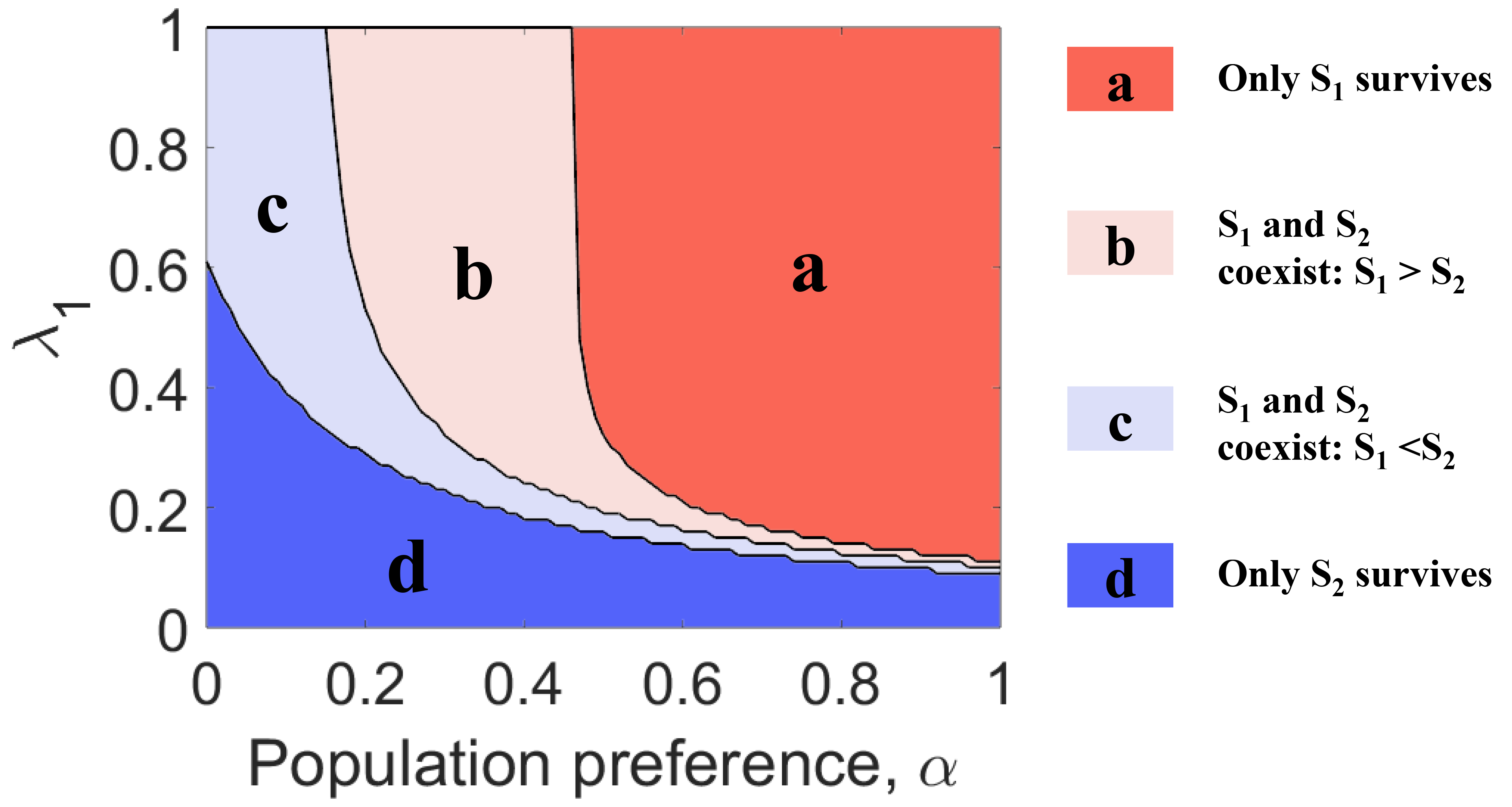}
\end{indented}
\caption{Phase diagram for competing diffusion results under different values of population preference and diffusion advantage. We fix $\lambda_2=0.2$, $\mu=0.2$. The phase plane is divided into four parts, represented by region (a)-(d), respectively: only $S_1$ survives, $S_1$ and $S_2$ coexist: $S_1\textgreater S_2$, $S_1$ and $S_2$ coexist: $S_1\textless S_2$, only $S_2$ survives. The separatrix lines are calculated numerically by equations (\ref{1})- (\ref{evolution equation}).}
\label{phase diagram}
\end{figure}

Furthermore, in figure \ref{phase diagram}, we explore the nonlinear joint effects of population preference and diffusion advantage. The phase plane is divided into four regions: (a) only $S_1$ survives, (b) $S_1$ and $S_2$ coexist: $S_1>S_2$, (c) $S_1$ and $S_2$ coexist: $S_1<S_2$, (d) only $S_2$ survives. All separatrix lines are calculated numerically by equations (\ref{1})- (\ref{evolution equation}). It is noteworthy that when $\lambda_1<\lambda_2=0.2$, i.e., information $2$ has the diffusion advantage, the competing results are possible to change from region (d) where only $S_2$ survives to region (a) where only $S_1$ survives as $\alpha$ varies from $0$ to $1$. This indicates the great power that the newly discussed population preference has on competing diffusion results. On the other hand, when $\alpha>0.5$, i.e., information $1$ has the population preference, the competing diffusion results are also possible to cross the four regions as $\lambda_1$ increases from $0$ to $1$.

\subsection{The effects of homophily}
In this section, we explore the detailed effects of homophily on competing diffusion results. We mainly focus on the following problems: How homophily influences diffusion advantage as well as population preference? In particular, can homophily reverse the evolutionary advantage, i.e., completely change the final competing diffusion results when other influencing factors are settled? Specifically, on what conditions the reversing phenomenon may happen?

\begin{figure}
\begin{indented}
\item[]\includegraphics[width=13cm]{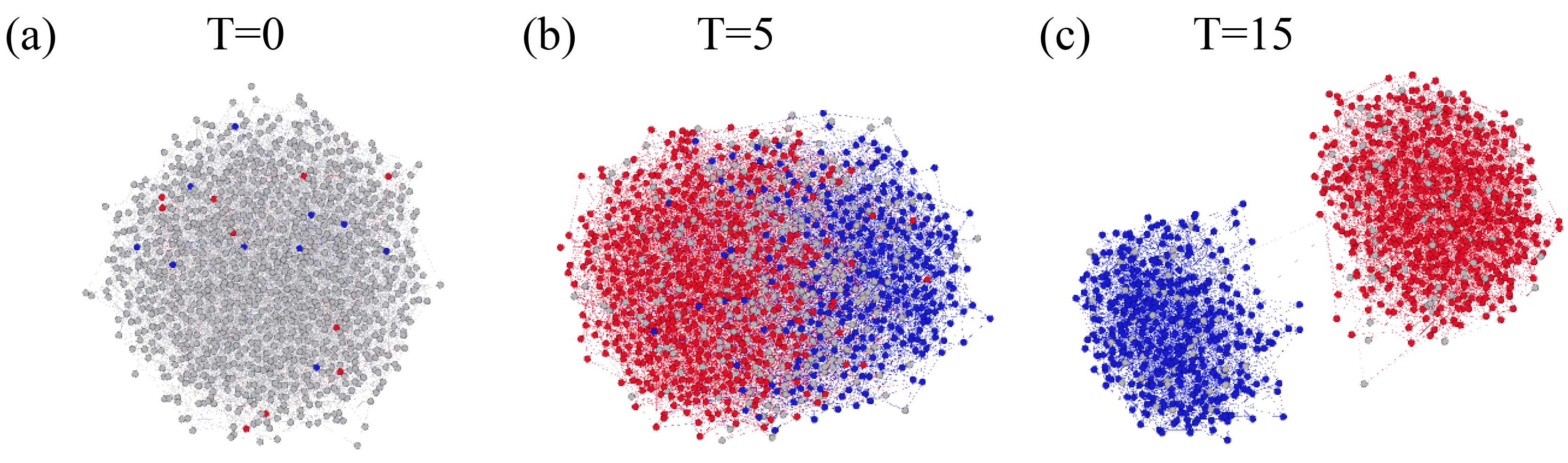}
\end{indented}
\caption{How homophily affects network structure over time. Spreader of information $1$ ($S_1$),  spreader of information $2$ ($S_2$) and ignorant are shown by red, blue and grey nodes, respectively. The parameters are set as follows: $\lambda_1=0.2$, $\lambda_2=0.15$, $\mu=0.1$, $\alpha=0$, $p=1$, $N=10^3$.  (a) $T=0$: $S_1=10$, $S_2=10$. (b) $T=5$: $S_1=490$, $S_2=284$. (c) $T=15$: $S_1=520$, $S_2=368$.}
\label{simulation}
\end{figure}

In figure \ref{simulation}, we show how network structure evolves when population homophily exists. The rewiring probability $p$ is set to be $1$. We find that the individuals gradually get close to those who spread the same information while leave away from those who support the different information as time goes on. At $T=15$, the whole network is completely divided into two clusters. Within each cluster, only one of the information survives, forming the echo chamber phenomenon which is widespread on social networks. Therefore, the population homophily can significantly affect the network structure and give rise to the formation of echo chambers, which will further influence the information diffusion paths and change the competing results.

\begin{figure}
\begin{indented}
\item[]\includegraphics[width=13cm]{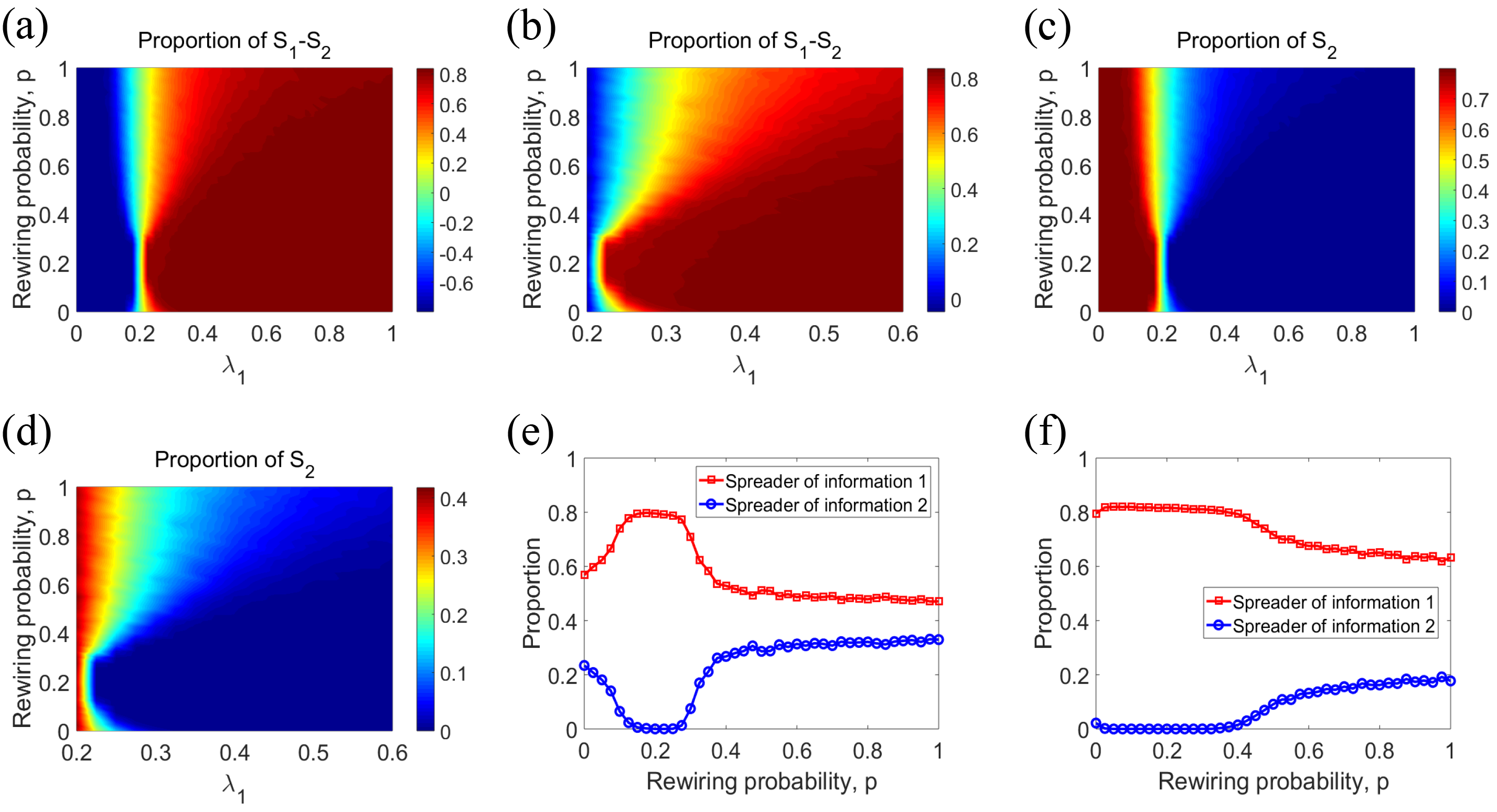}
\end{indented}
\caption{How homophily affects diffusion advantage in competing diffusion process. We fix $\alpha=0.5$, $\lambda_2=0.2$, $\mu=0.2$. (a) The evolutionary advantage, i.e. the final proportion of $S_1-S_2$, under different combinations of $\lambda_1$ and rewiring probability $p$. (b) A detailed view of (a) where $\lambda_1$ changes from $0.2$ to $0.6$. (c) The final proportion of $S_2$ under different combinations of $\lambda_1$ and rewiring probability $p$. (d) A detailed view of (c) where $\lambda_1$ varies from $0.2$ to 0.6. (e)(f) How homophily affects the effects of diffusion advantage. The final proportion of $S_1$ and $S_2$ are presented as a function of rewiring probability. We set $\lambda_1=0.225$ in (e) and $\lambda_1=0.3$ in (f), respectively. }
\label{transmission}
\end{figure}

We then study how homophily affects diffusion advantage in figure \ref{transmission}. We fix $\alpha= 0.5$ to exclude the influence of population preference. In figure \ref{transmission}(a), the evolutionary advantage, i.e. the final proportion of $S_1-S_2$, is presented under different combinations of $\lambda_1$ and rewiring probability $p$. Results show that the population homophily can change the evolutionary advantage arisen by diffusion advantage to some extent. When $\lambda_1$ is slightly larger than $\lambda_2$, which is equal to $0.2$ in our simulations, the evolutionary advantage first increases and then decreases as the rewiring probability grows. When $\lambda_1$ is much larger than $\lambda_2$, the evolutionary advantage reduces as $p$ increases. Figure \ref{transmission}(b) provides a detailed view of figure \ref{transmission}(a), where $\lambda_1$ varies from $0.2$ to $0.6$. We find the proportion of $S_1-S_2$ is always larger than zero. This indicates that without population preference, the population homophily could not reverse the evolutionary advantage caused by diffusion advantage. In figure \ref{transmission}(c) and (d), we show the global and local phase diagram of the proportion of $S_2$. As $p$ becomes larger, the critical value of $\lambda_1$ which makes $S_2$ extinct exhibits a slow decrease and then a rapid increase, indicating the fact that higher population homophily helps the information with lower transmission probability survive in the competition while lower homophily accelerates its extinction process. In figure \ref{transmission}(e) and (f), we give two detailed examples of how homophily affects the effects of diffusion advantage. The proportion of $S_1$ and $S_2$ are presented as a function of rewiring probability. We set $\lambda_1=0.225, 0.3$, respectively. To sum up, low homophily enhances diffusion advantage in competitive information diffusion while high homophily reduces it.

\begin{figure}
\begin{indented}
\item[]\includegraphics[width=13cm]{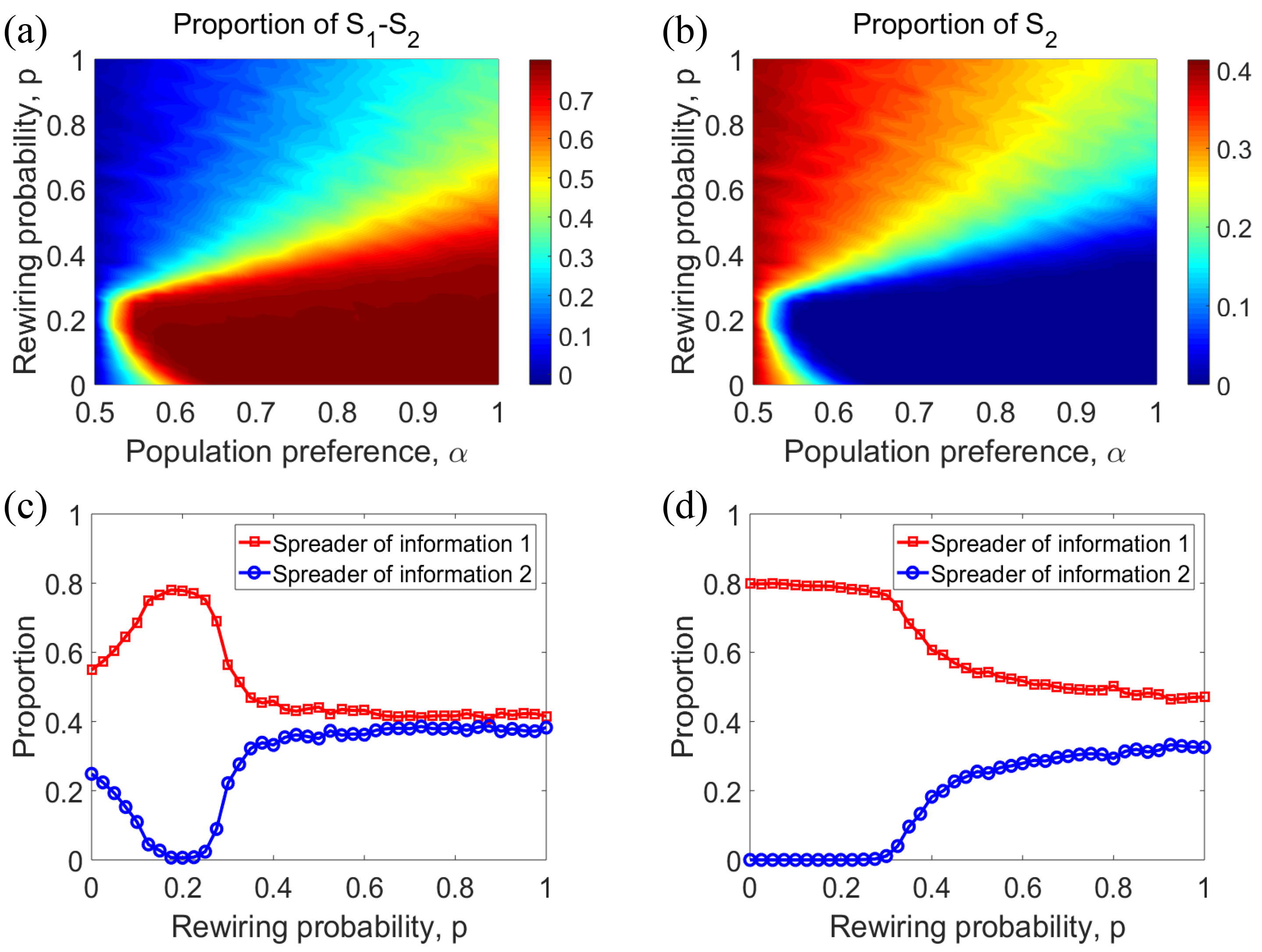}
\end{indented}
\caption{How homophily affects population preference in competitive information diffusion. We fix $\lambda_1=\lambda_2=0.2$, $\mu=0.2$. (a) The evolutionary advantage, i.e. the final proportion of $S_1-S_2$, under different combinations of population preference $\alpha$ and rewiring probability $p$. (b) The proportion of $S_2$ under different combinations of population preference $\alpha$ and rewiring probability $p$. (c)(d) How homophily affects the effect of population preference. The proportion of $S_1$ and $S_2$ are presented as a function of rewiring probability. We set $\alpha=0.55, 0.7$, respectively.}
\label{preference}
\end{figure}

Next, we discuss the impact of homophily on population preference in figure \ref{preference}. We set $\lambda_1=\lambda_2=0.2$ to exclude the effects of diffusion advantage. Due to the symmetry, we assume $\alpha\in[0.5, 1]$. In figure \ref{preference}(a), we present the evolutionary advantage under different combinations of population preference $\alpha$ and rewiring probability $p$. When $\alpha$ is slightly larger than $0.5$, the evolutionary advantage first rises and then declines as $p$ becomes larger. When $\alpha$ is much larger than 0.5, the evolutionary advantage reduces as $p$ increases. While the homophily does change the evolutionary advantage, the proportion of $S_1-S_2$ is always larger than zero. This indicates that without the existence of diffusion advantage, the evolutionary advantage solely caused by population preference can not be reversed by homophily effects. Figure \ref{preference}(b) further presents the phase diagram for the proportion of $S_2$. The critical value of $\alpha$ which results in the extinction of $S_2$ displays a gradual decrease followed by a sharp increase as the rewiring probability grows, which reveals that lower homophily accelerates the extinction of information with lower population preference while higher homophily helps it survive from the competition. In figure \ref{preference}(c) and (d), two typical examples of how homophily affects the effect of population preference are given. We set $\alpha=0.55, 0.7$, respectively. Results verify that lower homophily promotes population preference effects while higher homophily suppresses it.

Through observing the significant variations of evolutionary advantage, we have examined the detailed effects of homophily on diffusion advantage and population preference, respectively. For a better understanding of the simulation results, here we provide further explanations based on the dynamical model mechanisms. In general, the population homophily, which is modeled as the rewiring process, plays two main roles in competitive information diffusion processes. One is to advance people to avoid ineffective communications with opposite-minded individuals and to diffuse their information to ignorants or like-minded agents, which improves the efficiency of information diffusion and thus enhances the evolutionary advantage.  The other one is to form clusters of like-minded individuals, which separates the competing information into different communities and hence protects the information in disadvantage. Basically, the competing diffusion result is the balance of these two impacts of population homophily. When $p$ is small, the clustering speed is relatively slow compared to the information diffusion. In this situation, the role of improving the efficiency of communication dominates, which promotes the evolutionary advantage. On the other hand, when $p$ is large, the role of clustering takes the domination which facilitates the formation of two echo chambers like in figure \ref{simulation}, helping the disadvantaged information survives. Under this circumstance, the evolutionary advantage is reduced. However, for now, the evolutionary advantage can not be reversed by homophily when it is caused by a single competing factor, regardless of diffusion advantage or population preference. Naturally, we then raise an interesting question which leads to more complicated situations: if the evolutionary advantage emerges under the circumstances where the initial winning information owns population preference but has no diffusion advantage, or it takes diffusion advantage but loses population preference, can population homophily reverse the competing diffusion results?

\begin{figure}
\begin{indented}
\item[]\includegraphics[width=13cm]{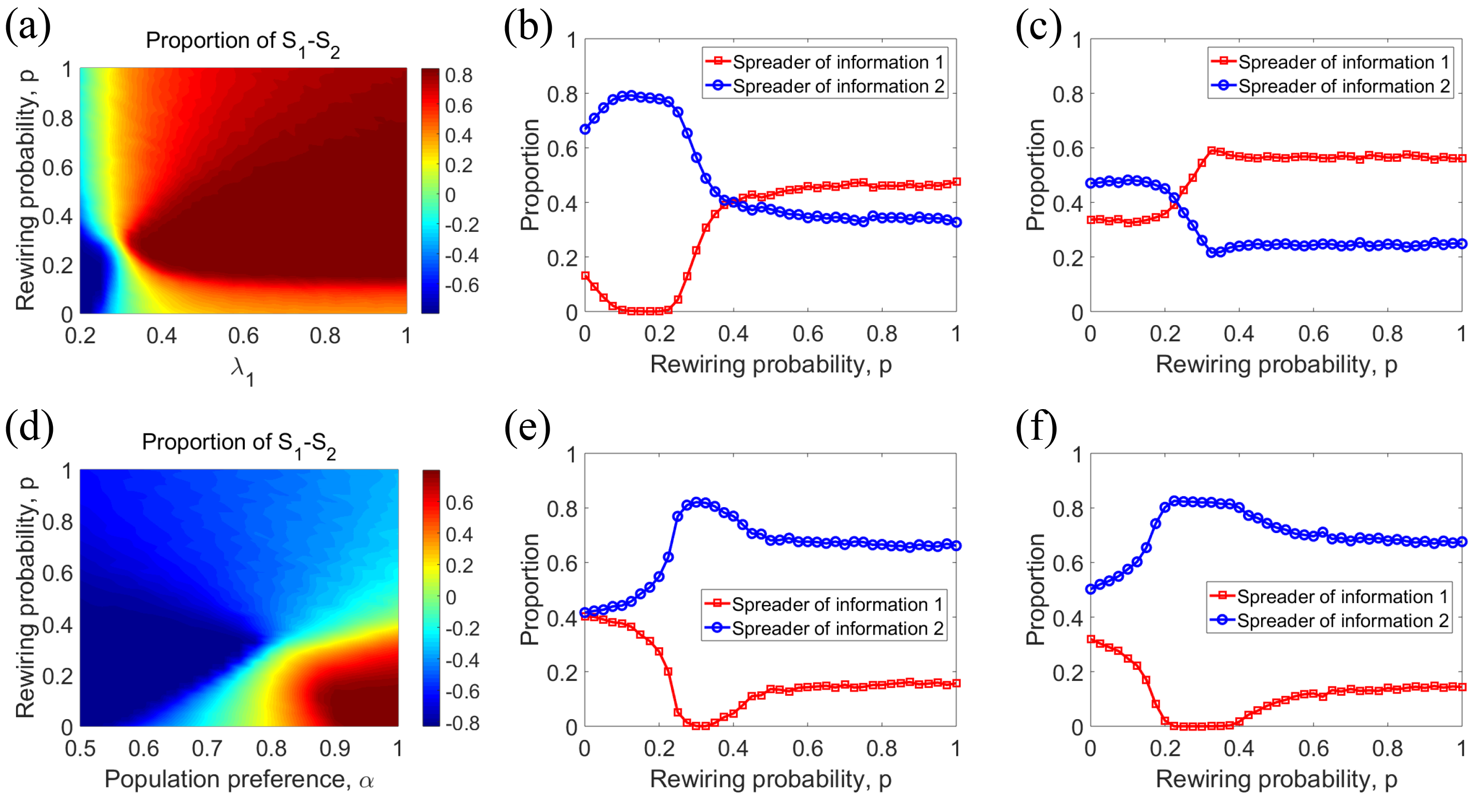}
\end{indented}
\caption{Homophily reverses the evolutionary advantage in competitive information diffusion on certain conditions. (a)-(c) The circumstance where information $1$ takes diffusion advantage while loses population preference. We set $\lambda_2=0.2$, $\mu=0.2$, $\alpha=0.3$. (a) The evolutionary advantage, i.e. the final proportion of $S_1-S_2$,  under different combinations of $\lambda_1$ and rewiring probability $p$. (b)(c) Two detailed reversing examples. We present the proportions of $S_1$ and $S_2$ as a function of rewiring probability. $\lambda_1=0.25, 0.3$, respectively. (d)-(f) The circumstance where information $1$ owns population preference but has no diffusion advantage. We fix $\lambda_1=0.2$, $\lambda_2=0.4$, $\mu=0.2$. (d) The evolutionary advantage under different combinations of population preference $\alpha$ and rewiring probability $p$. (e)(f) The proportions of $S_1$ and $S_2$ as a function of rewiring probability. $\alpha=0.745, 0.7$, respectively. }
\label{combination}
\end{figure}

In view of this question, in figure \ref{combination}, we further explore the reversing conditions in competitive information diffusion. First, in figure \ref{combination}(a)-(c), we examine the situation where information $1$ takes diffusion advantage while loses population preference. We fix $\lambda_2=0.2$, $\mu=0.2$, $\alpha=0.3$ and change $\lambda_1$ from $0.2$ to $1$. Figure \ref{combination}(a) shows that homophily can reverse the competing diffusion results, i.e., make $S_1-S_2$ change from negative to positive for a range of $\lambda_1$: $\lambda_1 \in [0.227, 0.33]$. To give a clear sight of this reversing phenomenon, in figure \ref{combination}(b) and (c), we show the variations of $S_1$ and $S_2$ as a function of rewiring probability $p$ in two detailed examples, where $\lambda_1=0.25, 0.3$ respectively. Results show that as $p$ increases, information $1$ goes through losing to winning, becoming stablely advantaged when $p$ is large. Then in figure \ref{combination}(d)-(f), we study the circumstance where information $1$ owns population preference but has no diffusion advantage. In figure \ref{combination}(d), when $\alpha>0.745$, $S_1-S_2>0$ initially which corresponds to the reversing situation discussed in figure \ref{combination}(a)-(c). Therefore here we focus on $\alpha<0.745$ in which situation the initially winning information (information 2) has diffusion advantage but loses population preference. Results show that the reversing phenomenon does not happen any more and information $2$ always wins as $p$ increases. Two typical examples are shown in figure \ref{combination}(e) and (f), where $\alpha=0.745$ and $0.7$, respectively. More parameter combinations are verified in {\it{Appendix A}} for this circumstance and no reversals are found. To sum up, population homophily is possible to reverse the evolutionary advantage in competitive information diffusion, but only if the initial winning information owns population preference but has no diffusion advantage. In other words, the only chance for the disadvantaged information to win is to occupy the diffusion advantage on large-scale social networks, regardless of the population preference.

\section{Conclusions and Discussions}\label{conclusions}
Competitive information diffusion is ubiquitous on online social networks, which directly influences the formation of public beliefs. For a better understanding of many real-world challenges such as rumor contagions and political social polarization, great effects have been made in exploring the underlying dynamical mechanisms for competing diffusion processes in recent years\cite{burghardt2016competing}. However, there still lacks a proper understanding of how population homophily, which is regarded as one of the most important collective human behaviors on large-scale social networks, affects the diffusion results of competitive information \cite{gu2014research}.

In this work, we propose a modified competitive SIS model, which incorporates homophily and generalized population preference to describe the simultaneous spread of two pieces of competitive information on large-scale social networks. Firstly, we examine how the modified model behaves without homophily. Using microscopic Markov Chain approach, we derive the phase diagram of competing diffusion results to show the impacts of diffusion advantage (i.e., the difference of transmission abilities of competitive information) and population preference. Then we explore the detailed effects of homophily, which is characterized by a rewiring mechanism. When population homophily is strong, the network structure evolves over time and finally forms two divided clusters, i.e., echo chambers, within which only one of the information survives. Results show that homophily can significantly change the evolutionary advantage. Lower homophily accelerates the extinction of disadvantaged information which enhances the evolutionary advantage, while higher homophily helps the disadvantaged information survive from the competition which reduces the evolutionary advantage. Further, we highlight the conclusion that homophily can even reverse the evolutionary advantage, but only when the initially disadvantaged information takes diffusion advantage while loses population preference. This indicates that the best chance for the information to win the competing evolution, either to secure the advantage or to reverse the disadvantage, is to occupy the diffusion advantage regardless of population preference.

Our work shows how population homophily makes influence on competitive information diffusion, which provides important insight into misinformation spreading, public opinion formation and many other competing dynamical processes on social networks. Moreover, the reversing condition based on the theoretical framework sheds light on designing effective competing strategies in a series of real situations, such as spreading new ideas, promoting industrial products and doing marketing. Our model also paves ways for further empirical studies in related scenarios.

\ack{This work is supported by Program of National Natural Science Foundation of China Grant No. 11871004.}

\begin{appendix}
\section{complementary studies for reversing conditions}
\begin{figure}
\begin{indented}
\item[]\includegraphics[width=13cm]{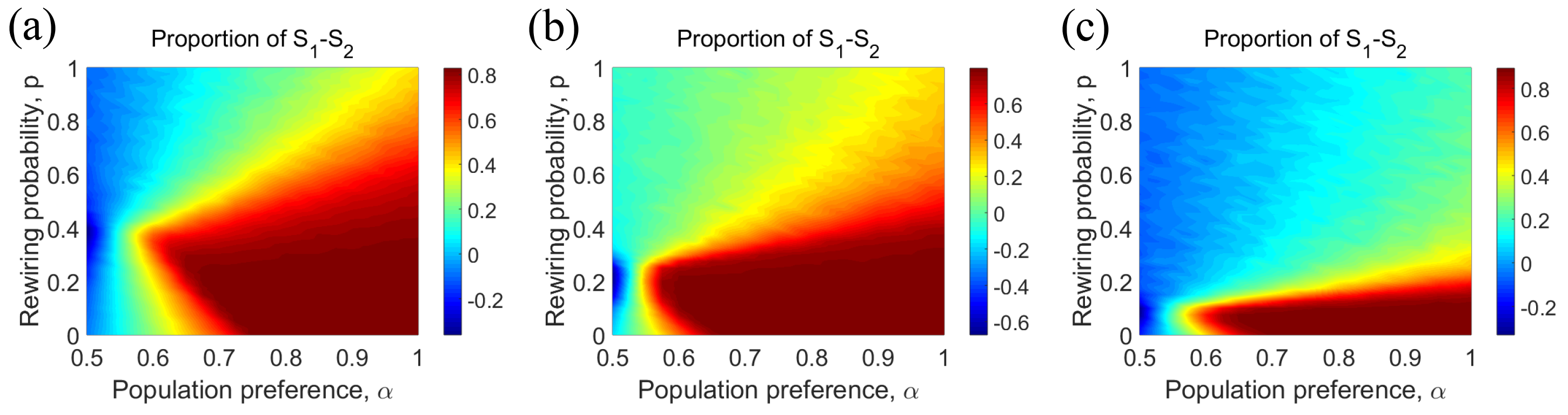}
\end{indented}
\caption{Exploring whether homophily can reverse the competing diffusion results when the initial winning information takes diffusion advantage but has no population preference. We present phase graphs of evolutionary advantage under different parameter combinations: (a) $\mu=0.2$, $\lambda_1=0.36$, $\lambda_2=0.4$; (b) $\mu=0.2$, $\lambda_1=0.19$, $\lambda_2=0.2$; (c) $\mu=0.1$, $\lambda_1=0.19$, $\lambda_2=0.2$.}
\label{fig8}
\end{figure}
In figure \ref{fig8}, we explore different parameter combinations to check whether reversals may happen under the circumstance that the initial winning information takes diffusion advantage but has no population preference. In figure \ref{fig8}(a) and (b), we fix $\mu=0.2$ and study the situations where the difference between transmission probabilities of the  competitive information (i.e., $\lambda_2 - \lambda_1$) is small, which are necessary complements for the results in figure \ref{combination}(d). Furthermore, in figure \ref{fig8}(c), we change $\mu$ to $0.1$ to check the influence of recovery rate in competing diffusion processes. Results show that no reversal happens when the initial winning information takes diffusion advantage, regardless of the population preference.

\end{appendix}

\providecommand{\newblock}{}

\end{document}